\documentclass[journal]{IEEEtran}

\usepackage{nicematrix}
\usepackage{color}
\usepackage[thinlines,thiklines]{easymat} 
\usepackage{mdwlist} 
\usepackage{amsmath,amsfonts,amssymb}
\usepackage{indentfirst}
\usepackage{graphicx}
\usepackage{upgreek}
\usepackage{subeqnarray}

\usepackage{amsmath}
\usepackage{graphicx}
\usepackage{amssymb}
\usepackage[latin1]{inputenc}
\usepackage{tikz}
\usetikzlibrary{positioning,calc}
\usepackage{arydshln}
\usepackage{mathdots}
\usepackage[normalem]{ulem}
\usepackage{tikz}
\usetikzlibrary{positioning,calc}
\usepackage{arydshln}
\usepackage{mathdots}
\usepackage[normalem]{ulem}
\usetikzlibrary{shapes,arrows}
\usetikzlibrary{matrix}
\usepackage{ifthen}
\usetikzlibrary{matrix,arrows,decorations.pathmorphing}
\usepackage[all]{xy}
\usepackage{bbold}
\usepackage{algorithm}
\usepackage{algpseudocode}
\usepackage{optidef}
\usepackage[percent]{overpic}
\usepackage{epstopdf}
\usepackage{rotating}
\usepackage{algorithm}
\usepackage{algpseudocode}
\usepackage{optidef}
\usepackage{mathrsfs}  
\usepackage{xcolor}

\algnewcommand\algorithmicforeach{\textbf{for each}}
\algdef{S}[FOR]{ForEach}[1]{\algorithmicforeach\ #1\ \algorithmicdo}

\renewcommand{\QED}{\QEDopen}

\newtheorem{definition}{Definition}

\allowdisplaybreaks
\makecompactlist{enumerateb}{enumerate}
\makecompactlist{itemizeb}{itemize}
\makecompactlist{descriptionb}{description}

\allowdisplaybreaks 

\def\aujour{\number\day \space \ifcase\month\or
janvier\or fvrier\or mars\or avril\or mai\or
juin\or juillet\or aot\or septembre\or octobre\or
novembre\or dcembre\fi \space \number\year}

\def\cH{{\cal H}}

\def\cL{{\cal L}}

\newtheorem{thm}{Theorem}
\newtheorem{lemma}{Lemma}
\newtheorem{cor}{Corollary}

\newtheorem{remark}{Remark} 

\def\C{{\setbox0=\hbox{$\displaystyle{\rm C}$}
		\hbox{\hbox to0pt{\kern 0.4\wd0\vrule height 0.95\ht0\hss}\box0}}}
\def\Q{{\setbox0=\hbox{$\displaystyle{\rm Q}$}%
	\hbox{\raise 0.2\ht0\hbox to0pt{\kern 0.4\wd0\vrule height 
	0.85\ht0\hss}\box0}}} 
\def\R{\mathop{\rm I\mkern -3.5mu R}} 

\def\cH2{{\cal H}_2} 

\def\cL2{\mathop{\mathcal L}_{2}} 

\def\cRH2{\mathop{\cal R \cal H}_2} 
\def\cRL2{\mathop{\cal R \cal L}_{2}} 

\DeclareMathOperator*{\rank}{rank}

\makeatletter 
\DeclareRobustCommand\sfrac[1]{\@ifnextchar/{\@sfrac{#1}}
                                            {\@sfrac{#1}/}}
\def\@sfrac#1/#2{\leavevmode\kern.1em\raise.5ex
         \hbox{$\m@th\fontsize\sf@size\z@
                           \selectfont#1$}\kern-.1em
         /\kern-.15em\lower.25ex
          \hbox{$\m@th\fontsize\sf@size\z@
                            \selectfont#2$}}
\makeatother

\begin{document}

\parindent=1.25 em

\title{Existence and Design of Target Output Controllers} 
\smallskip
\author{Tyrone Fernando and Mohamed Darouach
\thanks{T. Fernando is with the Department of Electrical, Electronic and Computer Engineering, University of Western Australia (UWA), 35 Stirling Highway, Crawley, WA 6009, Australia.}
\thanks{M. Darouach is with the University de Lorraine, Centre de Recherche en Automatique de Nancy (CRAN UMR-7039, CNRS), IUT de Longwy, 186 rue de Lorraine 54400, Cosnes et Romain, France. (email:  tyrone.fernando@uwa.edu.au, mohamed.darouach@univ-lorraine.fr)}
\thanks {The work of the second author is supported by Gledden Senior Visiting Fellowship, Institute of Advanced Studies (IAS), UWA, Australia.}}
%

\maketitle
\begin{abstract}                          
This paper introduces new conditions for target output controllability and provides existence conditions for placing a specific number of poles with a target output controller. Additionally, an algorithm is presented for the design of a target output controller. Controllability of the system under consideration is not required for designing target output controllers in this context. The findings in this paper extend the principles of full state feedback control. Moreover, we present conditions for static output feedback control under specific constraints. Several numerical examples are provided to illustrate the results.
\end{abstract}
\section{Introduction} 
Target output controllability is the generalised counterpart of controllability. Whilst controllability is about steering all the individual states of a system from any initial state to any desired state in finite time, target output controllability is about steering some linear functions of the states of the system from any initial state to any desired state in finite time. Uncontrollable systems are not unusual, but even in such systems, not all states are necessarily uncontrollable; some states or their linear combinations may still be influenced. Examples of systems with partially controllable state vectors are well-documented in the literature. For instance, in Chapter 5 of \cite{1aa}, where the ability to control a function of the states of a two-cart mechanical system is discussed, despite the inability to control its centre of mass. Similarly, Chapter 5 also explores an electrically balanced bridge circuit, where the voltage across a resistor cannot be controlled, yet some functions of the states are controllable. The notion of target output controllability plays a key role in designing controllers for systems where all the states are not controllable. 

Let us consider the following multivariable system $(A,B,C)$ as follows,
\begin{subeqnarray}\label{Rotella}
	\dot{x}(t) &=& Ax(t)+Bu(t), \slabel{eq1a}\\
	y(t) &=& Cx(t), \slabel{eq1b}
\end{subeqnarray}
with the initial state $x(0)=x_0$,  $x(t)\in \R^n$ is the state  vector, $u(t)\in \R^m$ is the known input, and $y(t)\in \R^p$ is the measurement output. Matrix $A\in \R^{n\times n}$, matrix $B\in \R^{n\times m}$ and matrix $C\in \R^{p\times n}$ is full row rank. The target output to be controlled is,
\begin{equation}\label{Rotella2}
	z(t)=Fx(t),
\end{equation}
where $F\in \R^{r\times n}$ is full row rank. Matrix $F$ is taken as a full row rank matrix because the linear functions to be controlled are all linearly independent, ensuring that any linearly dependent functions are also controlled. In literature when $F$ is chosen as $F=C$ then the problem is referred to as output controllability. However, $F$ need not necessarily be chosen as $F=C$, and for the general case the problem is referred to as target controllability or target output controllability.  For the special case when $F=I_n$, the target output controllability problem reduces to the classical full state controllability problem.

The problem to be addressed in the paper is to find a target output controller that will steer the target outputs from any initial state $z(t_0)$ to any desired state $z^{*}(t)$, where $t>t_0$. In particular, for $F \neq I_n$, this control objective is to be achieved without having to steer all the individual states from initial state $x(t_0)$ to some desired state $x^{*}(t)$. The benefit of such an approach of focusing on the target outputs alone is, it eliminates the controllability  requirement, in fact, the requirement is satisfaction of the more relaxed condition of target output controllability. Condition for output controllability was first proposed by Bertram and Sarachik in \cite{1a} where an algebraic rank condition for output controllability is reported, which generalises the well known  Kalman's  controllability test. In  \cite{5a}-\cite{14b}, further investigations on output controllability and its application to network control systems can be found.
 For investigations on functional observability see \cite{15a}-\cite{mdtfnew}, and functional observers,  see \cite{22new}-\cite{23b} and 
 for duality between output controllability and functional observability, see \cite{3a}, where it reports that the duality of the two concepts is not straightforward. It is only recently that a generalised PBH type rank condition for output controllability  was reported in \cite{5a}. However, in \cite{2a} a counter example is presented to demonstrate that the reported condition in  \cite{5a} is only necessary for output controllability. Moreover, in \cite{2a} a class of systems under which a generalised PBH test is sufficient and necessary for output controllability is reported. In this paper, we report a new condition to test if a system is target output controllable which corrects the target output controllability condition reported in \cite{5a}. In the numerical example section we show that the presented counter example in \cite{2a} correctly verifies the target output controllability of the system based on the new condition in this paper, consistent with the condition reported in \cite{1a}. Moreover, we present conditions for the existence of target output controllers for the placement of specific number of poles. We also present a target output controller  design algorithm. In the numerical examples section we consider uncontrollable systems and show the design of controllers. 

The paper is organised as follows: in section II we present the new criteria for target output controllability. Section III, reports the existence conditions and design of target output controllers by placement of $r$ system poles where $r$ is the number of target outputs to be controlled. We also present conditions for output feedback control under specific constraints. On the other hand, section IV, reports target output controller design by placement of $n_0$ system poles where $r<n_0\leq n$. Sections V and VI present numerical examples and conclusions respectively. We shall use the following notation, $X^T$ is the transpose of $X$, $X^-$ to denote the generalised inverse of matrix $X$ satisfying only the condition $XX^-X=X$, $\mathrm{eig}(X)$ is the set of eigenvalues of matrix $X$, $\mathcal{N}(X)$ 
represents a matrix whose columns form a basis for the null space of $X$, i.e., $X\mathcal{N}(X)=\bf{0}$, $\mathrm{rank}(X)$ is the rank of matrix $X$, $\mathcal{R}(X)$ is the range of $X$, $\mathbb{N}$ is the set of natural numbers including 0 and union of two sets is denoted by $\cup$. If the dimension of identity matrix is obvious then we represent it as $I$, otherwise it is shown in the subscript  of $I$.

\section{Criteria for Target Output Controllability}
Let us consider the linear time invariant system \eqref{eq1a}-\eqref{eq1b}.
The solution of \eqref{eq1a} with initial value $x(t_0)$ is given by,
\begin{equation}\label{eq-9}
	x(t)=e^{A(t-t_0)}x(t_0)+\int_{t_0}^{t} e^{A(t-\tau)}Bu(\tau)~d\tau, 
\end{equation}
and the target output at $t>t_0$, i.e., $z(t)$ is given by,
\begin{equation}\label{eq-9b}
	z(t)=Fx(t).
\end{equation}
Moreover, from \eqref{eq-9} and \eqref{eq-9b} we can also write the target output at $t > t_0$ with $x(t_0)=0$, i.e., $z_0(t)$, as follows,
\begin{equation}\label{eq-9c}
	z_0(t)=Fx(t)=F\int_{t_0}^{t} e^{A(t-\tau)}Bu(\tau)~d\tau,
\end{equation}
and we can also write the target output at $t>t_0$ with any $z(t_0)$, i.e., $z(t)$, as follows,
\begin{equation}
	z(t) = \tilde{z}(t) + z_0(t),
\end{equation}
where
\begin{equation}
\tilde{z}(t)= Fe^{A(t-t_0)}x(t_0).
\end{equation}
\noindent Before presenting the main results of this section, we shall give the following definitions and lemmas.\\
\begin{definition}\label{TFMDdef1}
	The {\it functional reachability} map on $[t_0, t]$ is defined to be,
	\begin{equation}\label{eqn7}
		L_{[t_0,t]}: u(\cdot) \mapsto F\int_{t_0}^{t} e^{A(t-\tau)}Bu(\tau)~d\tau.
	\end{equation}
\end{definition}

\begin{definition}
	The linear combinations of states $z(t)=Fx(t)$ of system \eqref{Rotella}, or the triple $(A,B,F)$, is target output controllable, if for any initial state $z(t_0)$ and any desired target state $z^{*}(t)$, there exists an input $u(t)$ that steers $z(t_0) = F x(t_0)$ to any desired target state $z^{*}(t) = F x(t)$ in finite time $t > t_0$.
\end{definition}
	
Using Cayley-Hamilton theorem we can express the term $A^n$ as follows,
$A^{n}=\sum_{k=0}^{n-1}\alpha_kA^{k}$, 
where $\alpha_k$ are the constants of the characteristic polynomial equation of order $n$, i.e., $f_n(\lambda)=0$, of system (1a),
\begin{equation}\label{TFMD5}
	f_n(\lambda)=\lambda^n - \sum_{k=0}^{n-1}\alpha_k\lambda^{k} = \mathrm{det}(\lambda I-A)=0.
\end{equation}
Considering Cayley-Hamilton theorem and the power series expansion of $e^{At}=I+At+\frac{1}{2!}A^2t^2+\hdots$,  lead to the following lemmas.
\begin{lemma}[\cite{24}]\label{lem-0}
	Given a characteristic polynomial equation of degree $n$, the exponential $e^{At}$ can be written as,
	\begin{equation}\label{TFMD6}
		e^{At} = \sum_{k=0}^{n-1}\beta_k(t)A^{k},
	\end{equation}
	where functions $\beta_k(t)$ are analytic.
\end{lemma}
\begin{lemma}
	The range of the functional reachability map, i.e., $\mathcal{R}(	L_{[t_0,t]})$  is, 
	\begin{equation}\label{TFMD7}
		\mathcal{R}(	L_{[t_0,t]})= \mathcal{R}(	F\mathcal{C}), 
	\end{equation}
	where  
	\begin{equation}\label{TFMD8}
		\mathcal{C}= \left(\begin{matrix}
			B &AB &\dots &A^{n-1}B
		\end{matrix}\right).
	\end{equation}
\end{lemma}
\begin{proof}
	Using \eqref{TFMD6} we can rewrite the following,
	\begin{IEEEeqnarray}{rcl}\label{TFMD9}
		&&F\int_{t_0}^{t} e^{A(t-\tau)}Bu(\tau)~d\tau,	\nonumber\\
		&=& F\int_{t_0}^{t} \sum_{k=0}^{n-1}\beta_k(t-\tau)A^{k}Bu(\tau)d\tau, \nonumber\\
		&=&F \sum_{k=0}^{n-1}A^{k}B\int_{t_0}^{t}\beta_k(t-\tau)u(\tau)d\tau. 
	\end{IEEEeqnarray}
	Using \eqref{TFMD8}, we can rewrite \eqref{TFMD9} as follows,
	\begin{equation}\label{tfmd15}
		F\int_{t_0}^{t} e^{A(t-\tau)}Bu(\tau)~d\tau = F\mathcal{C}c,
	\end{equation} 
	where $c=\left(\begin{matrix}
		c_0^T &\dots &c_{n-1}^T
	\end{matrix}\right)^T$ with $c_k=\int_{t_0}^{t}\beta_k(t-\tau)u(\tau)d\tau$ for $k\in \{0,\dots,n-1\}$.  
	From \eqref{tfmd15} and Definition \ref{TFMDdef1} we get \eqref{TFMD7}.
\end{proof}
\begin{lemma}\label{lm3}
	$L_{[t_0,t]}$ is surjective if and only if,
	\begin{equation}\label{TFMD13}
		\rank(F) =\rank(F\mathcal{C})=r.
	\end{equation}
\end{lemma}
\begin{proof}
	Since $x(t)\in \R^n$ and $z_0(t)=Fx(t)$ with $x(t_0)=0$, we obtain,
	\begin{equation}
		\mathcal{R}(z_0(t)) = \mathcal{R}(F).
	\end{equation}
	$L_{[t_0,t]}$ is surjective if and only if,
	\begin{equation}\label{TFMD15}
		\mathcal{R}(z_0(t))=\mathcal{R}\left(L_{[t_0,t]}\right),
	\end{equation}
	and from \eqref{TFMD7} and \eqref{TFMD15} we get \eqref{TFMD13}.
\end{proof}

Following theorem characterises target output controllability.
\begin{thm}\label{thm1MD}
	The following conditions are equivalent:
	\begin{enumerate}
		\item The linear combination of states $z(t)=Fx(t)$ of system \eqref{Rotella} is target output controllable or the triple $(A,B,F)$ is target output controllable.
		\item $\rank(F)=\rank(F\mathcal{C})$. 
		\item $\rank(F)=\\\rank \begin{pmatrix}FB &F(A-s I)B &\dots &F(A-s I)^{n-1}B\end{pmatrix}$, $\forall s \in \C$.
	\end{enumerate}
\end{thm}

Now let us define the following matrix before presenting the proof of Theorem 1: \\
\\
$\mathcal{P}=\begin{pmatrix}
	I&-s I&(-s)^2 I&\dots&(-s)^{n-1} I\\
	0&I&\binom{2}{1}(-s) I&\dots&\binom{n-1}{1}(-s)^{n-2} I\\
	0&0&\;\;I&\ddots&\vdots\\	
	\vdots&&\ddots&\;\;\;\;\;\;\ddots&\binom{n-1}{n-2}(-s) I\\
	0&\dots&\dots&0&\;\;I 
\end{pmatrix}$, 
where $\binom{n}{p}=\dfrac{n!}{(n-p)!p!}$.\\

	\begin{proof}
		%
		Notice that $L_{[t_0,t]}$ determines the set of target outputs $z(t)$ that can be reached from $x(t_0)=0$ at
		$t > t_0$. Suppose one wants to reach $z^{*}(t)$ at time $t >t_0$ from $z(t_0)$. From (6) and \eqref{eqn7} we get,
		\begin{equation}
			z^{*}(t) - \tilde{z}(t) = L_{[t_0,t] }u(\cdot).
		\end{equation}
		Therefore if $L_{[t_0,t]}$ is surjective, we can find $u(\cdot)$ to satisfy the above. Conversely, if the map is not surjective, we can pick $z^{*}(t)$ such that $(z^{*}(t)-\tilde{z}(t))$ cannot be reached by the map. Therefore from Lemma \ref{lm3} we obtain Condition $1) \Leftrightarrow$ Condition $2)$.
		On the other hand Condition $3)$ is equivalent to, 
		\begin{equation}
			\rank(F)=\rank(F\mathcal{C}\mathcal{P}), 
		\end{equation} 
		which proves, since matrix  $\mathcal{P}$ is non-singular, that condition $3)$ is equivalent to condition $2)$. 
	\end{proof}

From Theorem 1, we can give the following corollaries.
\begin{cor}
	If the pair $(A,B)$ is controllable then the triple $(A,B,F)$ is output controllable.
\end{cor}
\begin{proof}
If the pair $(A,B)$ is controllable then, $\rank(\mathcal{C})=n$. 
Since $\mathcal{C}$ is full row rank, we get $\rank(F) = \rank(F\mathcal{C}).$
\end{proof}	
\begin{cor}
If $F=I_n$, then the target output controllability condition 2) and also condition 3) of Theorem 1 reduce to the full state controllability condition of $\rank (\mathcal{C})=n$.
\end{cor}

%
\section{Existence Conditions and Target Output Controller Design by Placement of $r$ Poles}
 Target output controllability is a more relaxed condition than the controllability condition. In fact, controllability implies target output controllability, however, the reverse is not true. In the following we present a necessary condition for the triple $(A,B,F)$ to be target output controllable.

	\begin{thm}
		The necessary condition for the triple $(A,B,F)$ to be target output controllable is
		$\rank\begin{pmatrix} s F-FA&FB\end{pmatrix}=r, \forall s \in \C$.
	\end{thm}
	\begin{proof}
		The proof will be made by contradiction. Assume that $(A,B,F)$ is target output controllable and 
		$\rank\begin{pmatrix} s F-FA&FB\end{pmatrix}<r, s \in \C$, then there exists a vector $v\neq 0, v \in \R^{r}$ such that,
		\begin{equation}
			v^{T}\begin{pmatrix} s F-FA&FB\end{pmatrix}= {\bf 0}, s \in \C
		\end{equation}
		which implies that,
		\begin{equation}
			v^{T}FB = {\bf 0}, \label{MD1a}
		\end{equation}	
	and
	\begin{equation}
			s v^{T}F = v^{T}FA. \label{MD1}
	\end{equation}	
		By post-multiplying both sides of \eqref{MD1} by $B$ and by using \eqref{MD1a}, we obtain,
	\begin{equation}
		v^{T}FAB = {\bf 0}, 
	\end{equation}	
now by post-multiplying \eqref{MD1} by $AB$ we obtain,
\begin{equation}
v^{T}FA^{2}B = {\bf 0},
\end{equation}
and recursively we obtain,
			\begin{IEEEeqnarray}{rll}
		v^{T}FA^{3}B &=& {\bf 0}, \nonumber\\
		&\vdots&   \label{tfmd75}\\
		v^{T}FA^{n-1}B &=& {\bf 0},\nonumber
	\end{IEEEeqnarray}	
	which implies,
	\begin{equation}
		v^{T}F\begin{pmatrix}
			B &\hdots &A^{n-1}B
		\end{pmatrix} =  {\bf 0},
	\end{equation}	
	which is equivalent to, 
	\begin{equation} \label{neweqn47}
	\rank \left(F\begin{pmatrix}
		B &\hdots &A^{n-1}B
	\end{pmatrix} \right)<r,
\end{equation}
	and \eqref{neweqn47} contradicts that the triple $(A,B,F)$ is target output controllable. This proves the theorem.
\end{proof}

We will use the following lemmas in the sequel of the paper.		
			
\begin{lemma}\label{newlemma6}
	The following matrix $S$ where,
\begin{equation} \label{eqnS}
		S = \begin{pmatrix}
			F^{-} & I-F^{-}F &{\bf 0}\\
			{\bf 0} &{\bf 0} &I
		\end{pmatrix},
\end{equation}
is of full row rank.
\end{lemma}
\begin{proof}
	In fact, we have,
	\begin{IEEEeqnarray}{lll}
		&&\rank \begin{pmatrix}
			F^{-} &I-F^{-}F  &{\bf 0}\\
			{\bf 0} &{\bf 0} &I
		\end{pmatrix} \nonumber \\ 
		&&= \rank \begin{pmatrix}
			F^{-} &I-F^{-}F &{\bf 0}\\
			{\bf 0} &{\bf 0} &I 
		\end{pmatrix} \begin{pmatrix} I &F &{\bf 0} \\
			{\bf 0} &I &{\bf 0}\\
			{\bf 0} &{\bf 0} &I 
		\end{pmatrix}, \nonumber \\
		&&= \rank \begin{pmatrix}F^{-} &I &{\bf 0}\\
			{\bf 0} &{\bf 0} &I
		\end{pmatrix},
	\end{IEEEeqnarray}	
which proves the lemma.
\end{proof}

\begin{lemma}\label{newlemma7}
		If $\mathrm{rank}\begin{pmatrix}
			FA\\F
		\end{pmatrix} =\mathrm{rank}(F),$ then the following equality holds:
\begin{equation}
\rank\begin{pmatrix}
	sF-FA &FB
\end{pmatrix} = \rank\begin{pmatrix} sI  - FAF^- & FB\end{pmatrix}. 
\end{equation}
\end{lemma}
\begin{proof} 
	 First we can see that $\mathrm{rank}\begin{pmatrix}
	 	FA\\F
	 \end{pmatrix} =\mathrm{rank}(F)$ is equivalent to, 
	\begin{equation}
		\rank \left(\begin{pmatrix}
			FA\\F
		\end{pmatrix}\begin{pmatrix}
			F^- &I-F^-F
		\end{pmatrix}\right) = \rank\left(F\begin{pmatrix}
		F^- &I-F^-F
	\end{pmatrix}\right),
	\end{equation}
	or equivalently,
	\begin{equation}
		\rank \begin{pmatrix}
			FAF^-  &FA(I-F^-F)\\
			I_r &\bf{0}
		\end{pmatrix} = \rank\begin{pmatrix}
		I &\bf{0}
	\end{pmatrix},
	\end{equation}
	or equivalently,
	\begin{equation}
		FA(I-F^-F) = \mathbf{0}. \label{eqn41av2} 
	\end{equation}				
	Now since matrix $S$ according to \eqref{eqnS} is of  full row rank and $FA(I-F^-F) = \mathbf{0}$, we have, 
\begin{IEEEeqnarray}{rcl}
	&&\rank\begin{pmatrix}
		sF-FA &FB
	\end{pmatrix}  \nonumber\\	
	&&=\rank\begin{pmatrix}
		sF-FA &FB
	\end{pmatrix} \begin{pmatrix}
		F^- &I-F^-F &{\bf 0}\\
		{\bf 0} &{\bf 0} &I
	\end{pmatrix}, \nonumber \\	
	&&= \rank\begin{pmatrix} sI  - FAF^- & {\bf 0} & FB\end{pmatrix}, \nonumber\\
	&&=\rank\begin{pmatrix} sI  - FAF^- & FB\end{pmatrix},
\end{IEEEeqnarray}
which proves the lemma.
\end{proof}

\begin{lemma}\label{newlemma9}
	The following equation,
\begin{equation} \label{neweq52} 
MF-FA=0,
\end{equation} 
where $A \in \R^{n\times n}$, $F\in \R^{r\times n}$, $\mathrm{rank}(F)=r$ and $r \leq n$ are known matrices and 
$M \in \R^{r\times r}$ is an unknown matrix, has a solution
if and only if, 
\begin{equation} \label{neweqn54a}
	\mathrm{rank}\begin{pmatrix}
		FA\\F
	\end{pmatrix} =\mathrm{rank}(F),
\end{equation}
and in this situation $M$ satisfies,
\begin{equation}
	\mathrm{eig}(M) \subseteq \mathrm{eig}(A).
\end{equation}
\end{lemma}	
\begin{proof}
From \eqref{neweq52} we obtain the solution (see \cite{25}),
\begin{equation}
	M=FAF^-,
\end{equation}
if and only if \eqref{neweqn54a} is satisfied.
Since $F$ is a full row rank matrix, it follows $\begin{pmatrix}
	F\\
	\mathcal{N}(F)^T
\end{pmatrix}$ is a non-singular matrix and 
\begin{equation}
  \begin{pmatrix}
	F\\\mathcal{N}(F)^T
\end{pmatrix} \begin{pmatrix}
	F^-&\left(\mathcal{N}(F)^T\right)^-
\end{pmatrix}=\begin{pmatrix}
I &\mathbf{0}\\
\mathbf{0} &I
\end{pmatrix}.
\end{equation}
It now follows,
\begin{IEEEeqnarray}{rcl}
	\mathrm{eig}(A) &=& \mathrm{eig}\left(\begin{pmatrix}
		F\\\mathcal{N}(F)^T
	\end{pmatrix}A\begin{pmatrix}
	F^-&\left(\mathcal{N}(F)^T\right)^-
\end{pmatrix} \right), \nonumber\\
&=&\mathrm{eig} \begin{pmatrix}
	FAF^- &FA\left(\mathcal{N}(F)^T\right)^-\\
	\mathcal{N}(F)^TAF^- &\mathcal{N}(F)^TA\left(\mathcal{N}(F)^T\right)^-
\end{pmatrix}, \nonumber\\
&=&\mathrm{eig} \begin{pmatrix}
	FAF^- &MF\left(\mathcal{N}(F)^T\right)^-\\
	\mathcal{N}(F)^TAF^- &\mathcal{N}(F)^TA\left(\mathcal{N}(F)^T\right)^-
\end{pmatrix}, \nonumber\\
&=& \mathrm{eig} \begin{pmatrix}
	FAF^- &\mathbf{0}\\
	\mathcal{N}(F)^TAF^- &\mathcal{N}(F)^TA\left(\mathcal{N}(F)^T\right)^-
\end{pmatrix},\nonumber\\
&=& \mathrm{eig}\left(FAF^-\right)\cup \mathrm{eig}\left(\mathcal{N}(F)^TA\left(\mathcal{N}(F)^T\right)^-\right), \nonumber\\
&=& \mathrm{eig}\left(M\right)\cup \mathrm{eig}\left(\mathcal{N}(F)^TA\left(\mathcal{N}(F)^T\right)^-\right),
\end{IEEEeqnarray} 
which proves the lemma.
\end{proof}

\begin{remark}
	The generalised inverse matrix $G$, satisfying $FGF=F$ can be obtained from Theorem 2.4.1 of reference \cite{25}, in fact we have in general, 
	\begin{equation}
		G = F^-+U-F^-FUFF^-, \label{neweqn45}
	\end{equation}
	where $U$ is an arbitrary matrix of appropriate dimension. Equation \eqref{neweqn45} can also be rewritten, since $F$ is of full row rank, as follows,
	\begin{equation}
		G=F^- + (I-F^-F)U,
	\end{equation}
	where $F^-$ is any matrix satisfying $FF^-F=F$. For simplicity, and numerical computation, we can take $U={\bf 0}$ and $F^-$ as the Moore-Penrose inverse. One method to compute the Moore-Penrose inverse, $F^+$, is to use the singular values of $F$, i.e., in general,
	\begin{equation}
		F=U\begin{pmatrix}
			\Sigma &{\bf 0}\\
			{\bf 0} &{\bf 0}
		\end{pmatrix}V^T,
	\end{equation}
	and
	\begin{equation}
		F^+=V\begin{pmatrix}
			\Sigma^{-1} &{\bf 0}\\
			{\bf 0} &{\bf 0}
		\end{pmatrix}U^T,
	\end{equation}		
	where $U$ and $V$ are orthogonal matrices of appropriate dimensions, and $\Sigma$ is a diagonal matrix of appropriate dimensions with the singular values of $F$ on the diagonal. 
\end{remark}				

Pre-multiplying \eqref{eq1a} by $F$, we obtain,
\begin{equation} 
	F\dot{x}(t) = FAx(t) + FBu(t),
\end{equation}	
which can be written as,
\begin{IEEEeqnarray}{rcl}
	&&F\dot{x} = FA(I-F^-F+F^-F)x(t)+FBu(t), \nonumber\\
	&&= FAF^-Fx(t) + FA(I-F^-F)x(t)+FBu(t). \label{newtfmd62}
\end{IEEEeqnarray}

Now we are ready to give the following theorem.
	\begin{thm}
		The control law $u(t)=-ZFx(t)$, $Z\in \R^{m\times r}$ can drive the linear functional $Fx(t) \rightarrow \bf{0}$ as $t \rightarrow \infty$ at an arbitrary rate of convergence from any initial condition $Fx(t_0)$ by placing $r$ poles of a subsystem of order $r$ if and only if the following conditions are satisfied,
	\begin{IEEEeqnarray}{rcl}
		\rank\begin{pmatrix} 	
			FA\\F
		\end{pmatrix} &=& \rank(F), \IEEEyessubnumber \label{eqn36v2a}\\
			\rank\begin{pmatrix}
				sF-FA &FB
			\end{pmatrix} &=& \rank(F), \forall s\in \C. \IEEEyessubnumber \label{eqn36v2b}
		\end{IEEEeqnarray}
	\end{thm}
	\begin{proof}
		From Lemma 5 we know that $\rank\begin{pmatrix} 	
			FA\\F
		\end{pmatrix} =\rank(F)$ is equivalent to,
		\begin{equation}
			FA(I-F^-F) = \mathbf{0}. \label{48} 
		\end{equation}				
We can prove sufficiency as follows: if \eqref{eqn36v2a} or equivalently if \eqref{48} is satisfied then from \eqref{newtfmd62} we have the following subsystem of order $r$,
\begin{equation}
	F\dot{x}(t) = FAF^- Fx(t) + FB u(t). \label{mdeqn43}
\end{equation}
With $u(t)=-ZFx(t)$, equation \eqref{mdeqn43} becomes,
\begin{equation}
	F\dot{x}(t) = \left(FAF^--FBZ\right)Fx(t). \label{eqnmd47}
\end{equation}
It now follows that $Fx(t) \rightarrow \mathbf{0}$ as $t \rightarrow \infty$ at an arbitrary convergence rate by placing $r$ poles of the subsystem of order $r$ if and only if the pair  $(FAF^-, FB)$ is controllable, which is equivalent to $\rank \begin{pmatrix}sI-FAF^- &FB\end{pmatrix} = \rank(F) = r \,\,\forall\,\, s\in \C$, which is also equivalent to \eqref{eqn36v2b} from Lemma \ref{newlemma7}. 

We will use the contrapositive to prove necessity, in fact if \eqref{eqn36v2a} or \eqref{48} is satisfied and \eqref{eqn36v2b} is not satisfied or equivalently if \eqref{eqn36v2a} or \eqref{48} is satisfied and the pair $(FAF^-, FB)$ is not controllable then from \eqref{eqnmd47} it follows that $Fx(t) \not \rightarrow \mathbf{0}$ as $t \rightarrow \infty$ at an arbitrary convergence rate by placing $r$ poles of the subsystem of order $r$. This proves the theorem.
\end{proof}

			Now let $\Lambda$ be the set of specified $r$ eigenvalues anywhere on the left half of the complex plane,
			\begin{equation}
				\Lambda =\{\lambda_1,\dots,\lambda_r\}.
			\end{equation}		
			\begin{thm}
				If conditions \eqref{eqn36v2a} and \eqref{eqn36v2b} are satisfied, the control law $u(t)=-ZFx(t)$, $Z\in \R^{m\times r}$,  can place $r$ eigenvalues of $(FAF^{-}-FBZ)$ at $\Lambda$, and these eigenvalues are a subset of the eigenvalues of $(A-BZF)$.
			\end{thm}	  
			\begin{proof}
		From Theorem $3$ we can see that under condition \eqref{eqn36v2a} and $u(t)=-ZFx(t)$, we have \eqref{eqnmd47}. Under condition \eqref{eqn36v2b}, the pair $(FAF^{-},FB)$ is controllable, and there exists a matrix $Z$ such that $\mathrm{eig}(FAF^{-} -FBZ)=\Lambda$.
		Moreover, under condition \eqref{eqn36v2a}, $\left(FAF^{-}\right)F-FA=\mathbf{0}$ (see equation \eqref{48}), which can be rewritten as the following Sylvester equation,
		\begin{equation}
			(FAF^{-} -FBZ)F-F(A-BZF)=\mathbf{0}. \label{neweqn55}
		\end{equation}
 Since $F \neq \mathbf{0}$ is full row rank, from Lemma \ref{newlemma9} it follows that $\mathrm{eig}(FAF^{-} -FBZ) \subseteq \mathrm{eig}(A - BZF)$. This proves the theorem.
\end{proof}

Substituting $F=C$ in Theorem 3 and Theorem 4, we establish the following result. 
\begin{cor}
The static output feedback control law $u(t)=-Zy(t)$, $Z\in \R^{m\times p}$ can drive the 
	linear functional $y(t) \rightarrow \bf{0}$ as $t \rightarrow \infty$ at an arbitrary rate of convergence from any initial condition $y(t_0)$ by placing $p$ poles of a subsystem of order $p$ at any desired set $\Lambda  =\{\lambda_1,\dots,\lambda_p\},$ where $\lambda_i \in \C, i \in \{1,\dots,p\}$, 
 if and only if the following conditions are satisfied,
	\begin{IEEEeqnarray}{rcl}
	\rank\begin{pmatrix} 	
		CA\\C
	\end{pmatrix} &=& \rank(C), \IEEEyessubnumber \label{tyeqn52a}\\
	\rank\begin{pmatrix}
		sC-CA &CB
	\end{pmatrix} &=& \rank(C) = p, \forall s\in \C.    \IEEEyessubnumber \label{tyeqn52b}
\end{IEEEeqnarray}
The gain $Z$ is determined using a pole placement method such that  
$\mathrm{eig}\left(CAC^--CBZ\right) = \Lambda.$ Moreover,  $\Lambda$ is a subset of the eigenvalues of $A-BZC$.

\begin{remark}	
	By placing the poles of the pair $(CAC^-, CB)$ as per Corollary 3, we are effectively assigning
	$p$ eigenvalues of the closed loop system matrix $(A-BZC)$
	anywhere in the complex plane. From Lemma \ref{newlemma9}, the remaining eigenvalues of $(A-BZC)$ are at $\mathrm{eig}\left(\mathcal{N}(C)^TA\left(\mathcal{N}(C)^T\right)^-\right).$
	For illustration, consider the controllable but unobservable system $(A,B,C)$ where $A=\begin{pmatrix}
		2 &0 &0\\-1 &1 &0\\0 &1 &-1
	\end{pmatrix}$, $B=\begin{pmatrix}
	1 \\0\\0
\end{pmatrix}$, and $C=\begin{pmatrix}
1 &1 &0
\end{pmatrix}$. We place the pole of the pair  $(CAC^-, CB)$ at -3 and find $Z= 4$. We can see that $-3$ is also an eigenvalue of $(A-BZC)$ as well. The remaining eigenvalues of $(A-BZC)$ are at $\mathrm{eig}\left(\mathcal{N}(C)^TA\left(\mathcal{N}(C)^T\right)^-\right) = \{-1, 2\}$.
%
%
	\end{remark}

\end{cor}	

\section{Target Output Controller Design by Placement of $n_0$ Poles}
Even if the triple $(A,B,F)$ is target output controllable, conditions \eqref{eqn36v2a} may not be satisfied, which is one of the two requirements for designing a target output controller by placing $r$ poles of a subsystem of order $r$. However, we may still design a target output controller by placing $n_0$ poles, where $r<n_0 \leq n$, provided the triple $(A,B,F)$ is target output controllable. This section demonstrates this approach. We first show that conditions \eqref{eqn36v2a} and \eqref{eqn36v2b}  are sufficient for the triple $(A,B,F)$ to be target output controllable in the following theorem.
\begin{thm}
If \eqref{eqn36v2a} and \eqref{eqn36v2b} are satisfied, then the triple $(A,B,F)$ is target output controllable.  
\end{thm}
\begin{proof}
Since \eqref{eqn36v2a} is equivalent to $FA(I-F^-F) = \bf{0}$ (see equation \eqref{48}), we obtain, $FAF^-F = FA,$
and post-multiplying both sides with $B$, we obtain,
\begin{IEEEeqnarray}{rcl}
\big(FAF^-\big)FB &=& FAB,\label{mdeqn53}
\end{IEEEeqnarray}	
Pre-multiplying both sides of \eqref{mdeqn53} with $FAF^-$ and using $FAF^-F=FA$, we obtain,
\begin{equation}
	(FAF^-)^2FB = FA^2B. \label{mdeqn54}
\end{equation} 
Again pre-multiplying both sides of \eqref{mdeqn53} with $FAF^-$ and using $FAF^-F=FA$, and continuing on the pre-multiplication process, we obtain,
\begin{IEEEeqnarray}{rcl}
	(FAF^-)^3FB &=& FA^3B  \nonumber\\
&\vdots&   \label{mdeqn55}\\
\big(FAF^-\big)^{n-1}FB &=& FA^{n-1}B. \nonumber	
\end{IEEEeqnarray} 
Since \eqref{eqn36v2b} is equivalent to pair $(FAF^-,FB)$ is controllable, i.e.,
\begin{equation}
	\rank\begin{pmatrix}
		FB &(FAF^-)FB  &\dots &(FAF^-)^{n-1}FB 
	\end{pmatrix} = r, \label{tfmd76}
\end{equation}
 using \eqref{mdeqn53}-\eqref{mdeqn55} in \eqref{tfmd76} we obtain,
 \begin{equation}
 	\rank\begin{pmatrix}
 		FB &FAB  &\dots &FA^{n-1}B 
 	\end{pmatrix} = \rank(F)= r, \label{tfmd77}
 \end{equation}
which proves the theorem.
\end{proof}

			If \eqref{eqn36v2a} and \eqref{eqn36v2b} are not satisfied, then those conditions can be relaxed by introducing additional target outputs $Rx(t) \in \R^{n_0-r}$ in a new augmented full row rank target output matrix $
				\begin{pmatrix}
					F\\
					R
				\end{pmatrix}$ as per the following corollary.\\
			\begin{cor}
			The control law $u(t)=-Z\begin{pmatrix}F\\R\end{pmatrix}x(t)$, $Z\in \R^{m\times n_0}$ can drive the 
				linearly independent functions $\begin{pmatrix}F\\R\end{pmatrix}x(t) \rightarrow 0$ as $t \rightarrow \infty$ at an arbitrary rate of convergence from any initial condition $\begin{pmatrix}F\\R\end{pmatrix}x(t_0)$ by placing $n_0$ poles of a subsystem of order $n_0$ if and only if the following conditions are satisfied,
					\begin{IEEEeqnarray}{rcl}
						\rank\begin{pmatrix} 	
							FA\\RA\\F\\R
						\end{pmatrix} &=& \rank\begin{pmatrix}F\\R\end{pmatrix}, \IEEEyessubnumber \label{tfmdneweqn79}\\
					\rank\begin{pmatrix}
						sF-FA &FB\\sR-RA &RB
					\end{pmatrix} &=& \rank\begin{pmatrix}F\\R\end{pmatrix}, \forall s\in \C. \IEEEyessubnumber \label{eqn62v2b}
				\end{IEEEeqnarray} 
			\end{cor}
			\begin{proof}
				Replace $F$ with $\begin{pmatrix}F\\R\end{pmatrix}$ and $r$ with $n_0$ in the proof of Theorem 3.
			\end{proof}	

If condition \eqref{eqn36v2a} is not satisfied then matrix $R$ can be determined by considering the observability indices of the pair $(A,F)$. Let $\nu_1, \dots,\nu_r$ be the observability indices that correspond to rows of $F$, i.e., $F_1, \dots, F_r$ respectively where,
	\begin{equation}
		\begin{pmatrix}
			F_1\\
			\vdots\\
			F_r
		\end{pmatrix} = F.
	\end{equation}
\begin{thm}
Matrix $R$ that satisfies \eqref{tfmdneweqn79} is,
	\begin{equation}\label{tfmdneweqn55}
R=O=\begin{pmatrix}
	O_1\\
	\vdots\\
	O_r
\end{pmatrix},
 	\end{equation}
 where for $i=1$ to $i=r$,
 \begin{equation}
 	O_i=\begin{pmatrix}
 	F_{i}A\\
 	\vdots\\
 	F_{i}A^{\nu_i-1}
 	\end{pmatrix}.
 \end{equation}
\end{thm}
\begin{proof}
Since $\rank\begin{pmatrix}
	F_i\\O_i
\end{pmatrix}=\nu_i$, the right hand side of \eqref{tfmdneweqn79} can be written as,

\begin{equation}
	\rank\begin{pmatrix}
		F\\R
	\end{pmatrix} = \sum_{i=1}^{r} 
	\rank\begin{pmatrix}
		F_i\\O_i
	\end{pmatrix} = \sum_{i=1}^{r} \nu_i.
\end{equation}
Since $\nu_1, \dots,\nu_r$ are observability indices of the pair $(A,F)$, we can write the following for $i=1$ to $i=r$, 
\begin{equation}
	\rank\begin{pmatrix}
		F_iA^{\nu_i}\\F\\O
	\end{pmatrix} = \rank \begin{pmatrix}
	F\\O
\end{pmatrix},
\end{equation}
and the left hand side of \eqref{tfmdneweqn79}  can be written as,
 \begin{equation}
 	\rank\begin{pmatrix}
 		F_1A^{\nu_1}\\ \vdots\\F_rA^{\nu_r} \\F\\O
 	\end{pmatrix} = \rank \begin{pmatrix}
 		F\\O
 	\end{pmatrix} = \rank \begin{pmatrix}
 	F\\R
 \end{pmatrix}=\sum_{i=1}^{r} \nu_i,
\end{equation}
which proves the theorem.
\end{proof}		

Now let $\bar{\Lambda}$ be the set of specified $n_0$ eigenvalues anywhere on the complex plane,
		\begin{equation}
			\bar{\Lambda} =\{\lambda_1,\dots,\lambda_{n_0}\}.
		\end{equation}
	
		From Theorem 4, we can write the following corollary.
			\begin{cor}\label{cor2}
				If conditions \eqref{tfmdneweqn79} and \eqref{eqn62v2b} are satisfied, the control law $u(t)=-Z\begin{pmatrix}F\\R\end{pmatrix}x(t)$, $Z\in \R^{m\times n_0}$, can place $n_0$ eigenvalues of $\left(\begin{pmatrix} 	
						F\\R\end{pmatrix}A\begin{pmatrix} 	
						F\\R\end{pmatrix}^{-}-\begin{pmatrix} 	
						F\\R\end{pmatrix}BZ\right)$ at $\bar{\Lambda}$, and these eigenvalues are a subset of the eigenvalues of $\left(A-BZ\begin{pmatrix} 	
					F\\R\end{pmatrix}\right)$. 	
			\end{cor}
			\begin{proof}
				Replace $F$ with $\begin{pmatrix}F\\R\end{pmatrix}$ and $\Lambda$ with $\bar{\Lambda}$ in the proof of Theorem 4.	
			\end{proof}


We can now present the following target output controller design algorithm.
\\

\indent {\it Target Output Controller Design Algorithm}
 \begin{itemize}[ ]
    \item[1:] Check if the triple $(A, B, F)$ is Target Output Controllable using Theorem 1, if yes continue to step 2, otherwise stop.
\item[2:] Check if condition \eqref{eqn36v2a} is satisfied, if yes then continue to step 3, otherwise go to step 4.
    \item[3:] Check if condition \eqref{eqn36v2b} is satisfied or if the pair $(FAF^-, FB )$ is controllable, if no then exit algorithm, otherwise determine $u(t)$ according to $u(t)=-ZFx(t)$ where $Z$ is determined through a pole placement of the pair $(FAF^-, FB )$.  Exit algorithm.
    \item[4:] Determine $R$ according to \eqref{tfmdneweqn55} and continue to Step 5.
	\item[5:] Check if condition \eqref{eqn62v2b} is satisfied or if the pair $\left(\begin{pmatrix} 	
		F\\R\end{pmatrix}A\begin{pmatrix} 	
		F\\R\end{pmatrix}^{-}, \begin{pmatrix} 	
		F\\R\end{pmatrix}B\right)$ is controllable, if no then exit algorithm, otherwise 
	determine $u(t)$ according to $u(t)=-Z\begin{pmatrix}
		F\\R
	\end{pmatrix}x(t)$ where $Z$ is determined through a pole placement of the pair $\left(\begin{pmatrix} 	
	F\\R\end{pmatrix}A\begin{pmatrix} 	
	F\\R\end{pmatrix}^{-}, \begin{pmatrix} 	
	F\\R\end{pmatrix}B\right)$. Exit algorithm.
\end{itemize}

\section{Numerical Examples}
\noindent {\it Example 1:} 

Let us consider the example presented in \cite{2a}, which is a counter example for the target output controllability condition reported in   \cite{5a}, where the system matrices $A, B, F$ as follows,

$A=\begin{pmatrix}
	0 &1 &0\\
	0 &0 &1\\
	0 &0 &0
\end{pmatrix}
$, $B=\begin{pmatrix}
	1\\
	0 \\
	0 
\end{pmatrix}
$ and $F=\begin{pmatrix}
	0 &1 &0
\end{pmatrix}.
$
\\

\noindent Based on condition 2 of theorem 1 in this paper, which was first reported in \cite{1a}, the triple $(A,B,F)$ is not output controllable. However, based on the following equivalent condition for target output controllability reported in \cite{5a}, i.e., 
$$
\rank \big(\begin{array}{cc}
	F(s I - A) &FB
\end{array}\big) = \rank (F), \forall \,\, s \in\C,
$$
we find the above condition is satisfied for all $s \in \C$, therefore based on the above condition reported in \cite{5a}, the system is output controllable, which is incorrect. Based on the new condition reported in this paper, which is condition 3 of theorem 1, it is violated at $s=0$, therefore correctly determines the triple $(A, B, F)$ is not output controllable.\\

\noindent {\it Example 2:} \\
\indent This example illustrates the design of target output controllers by placement of $r$ poles. Let,\\

$
A = \left(\begin{array}{rrrrr}
 1    &0.5   &-1         &0    &1\\
0.3   &0.5   &-0.6   &-0.3    &0.3\\
-0.6        &0    &0.2    &0.6   &-0.6\\
1.25    &0.5   &-1   &-0.25    &1.75\\
-0.75         &0         &0    &0.75   &-0.25
\end{array}\right)
$,\\
$
B=\left( \begin{array}{rr}
1 & -1 \\
1 & 1 \\
0 & 0 \\
1 & 0 \\
0 & 1 
\end{array}\right)
$, $C = \begin{pmatrix}0 & 0 & 2 & 1 & 0 \\
0 & 0 & 0 & 0 & 1\end{pmatrix}$ and
$$F = \begin{pmatrix}1 &1 &-2 &0 &2\end{pmatrix}.$$ 

\noindent We now follow the target output controller design algorithm and first establish that the triple $(A, B, F)$ is target output controllable and verify that condition condition \eqref{eqn36v2a} is satisfied. Moreover, the pair $(FAF^-, FB)$ is controllable or equivalently, condition \eqref{eqn36v2b} is satisfied. Using pole placement we can determine,
$$Z=\begin{pmatrix}
	0.75\\
	0.75
\end{pmatrix},
$$ in order to place the pole of the pair $(FAF^-,FB)$ at -2. It is also noted that $A-BZF$ has an eigenvalue at -2 as per Theorem 4.\\

\noindent {\it Example 3:} \\
\indent This example illustrates the design of target output controllers by placement of $n_0$ poles. Let us now consider the same $A, B$ and $C$ matrices as in Example 2. Let matrix $F$ be as follows:
$$
F = \begin{pmatrix}0.5 &1 &-2 &0.5 &2.5\end{pmatrix}.
$$ 
We now follow the target output controller design algorithm and first establish that the triple $(A, B, F)$ is target output controllable and verify that condition \eqref{eqn36v2a} is not satisfied. We note that the observability index of $(A,F)$ is 2 and determine $R$  according to \eqref{tfmdneweqn55} as follows:\\
$$
R=FA= \begin{pmatrix}
0.75    &1  & -2   &0.25    &2.25
\end{pmatrix}.
$$ 
\\
\noindent We can verify that condition \eqref{tfmdneweqn79} is satisfied. Moreover, the pair $\left(\begin{pmatrix} 	
	F\\R\end{pmatrix}A\begin{pmatrix} 	
	F\\R\end{pmatrix}^{-}, \begin{pmatrix} 	
	F\\R\end{pmatrix}B\right)$ is controllable or equivalently, condition \eqref{eqn62v2b} is satisfied. Using pole placement we can determine, $$Z=\begin{pmatrix}
	-6.5    &11\\
	5   &-7
\end{pmatrix},$$ in order to place the poles of the pair $\left(\begin{pmatrix} 	
	F\\R\end{pmatrix}A\begin{pmatrix} 	
	F\\R\end{pmatrix}^{-}, \begin{pmatrix} 	
	F\\R\end{pmatrix}B\right)$ at -2 and -3. It is also noted that $A-BZF$ has eigenvalues at -2 and -3 as per Corollary 5.\\

\noindent {\it Example 4:} \\
\indent This example illustrates the design of a static output feedback controller by placement of $p=2$ poles anywhere in the complex plane of a second order subsystem to drive $y(t) \rightarrow \bf{0}$ as $t \rightarrow \infty$ from any initial value $y(t_0)$. The considered system is unstable, uncontrollable and unobservable where system matrices $A, B$ and $C$ as follows:\\
$$
	A = \begin{pmatrix}
	-0.5    &0.5  & -1   &-0.5   &0.5\\
	-0.7   &-0.5    &1.4    &0.7   &-0.7\\
	-0.6   &0    &0.2    &0.6   &-0.6\\
	0.25   &0.5   &-1   &-1.25   &0.75\\
	-0.25   &0  &0 &0.25 &-0.75
	\end{pmatrix},
$$ $B = \begin{pmatrix}1  &-1\\
2  &1\\
0.5  &1\\
1  &-1\\
0   &2
\end{pmatrix}$ and $C = \begin{pmatrix}
 0.5 &0 &0  &-0.5  &0.5\\
-0.5  &0  &2   &0.5 &-0.5
\end{pmatrix}.$ \\
\\

\noindent The eigenvalues of $A$ are at  0.2, -0.5, -1, -1 and -0.5.
We can verify that \eqref{tyeqn52a} and \eqref{tyeqn52b} are satisfied. Using the pole placement technique, we can determine, \\
$$Z=\begin{pmatrix}
	-2.2    &3.2\\
	1   &0
\end{pmatrix},$$ 
\\
\noindent in order to place the poles of the pair $\left(CAC^-, FB\right)$ at -2 and -3. We also note that the static output feedback control law $u(t)=-Zy(t)$ makes the closed loop system as follows,
\\
$$
\dot{x}(t)=Ax(t)-BZy(t)=(A-BZC)x(t),
$$
\\
\noindent and that the eigenvalues of $(A-BZC)$ include -2 and -3 which are also the eigenvalues of $\left(CAC^--FBZ\right)$. The remaining eigenvalues of $(A-BZC)$ are at $\left(\mathcal{N}(C)^TA\left(\mathcal{N}(C)^T\right)^-\right)$ which are -0.5, -0.5 and -1 as per Corollary 4. Also note that -0.5, -0.5 and -1 are the eigenvalues of $A$ which are unaltered. 

\section{Conclusion} 
A new criteria for testing target output controllability is presented. We have also presented target output controller existence conditions for placing specific number of poles, and presented a target output controller design algorithm. Moreover, we have presented conditions for static output feedback control by placement of $p$ poles of a subsystem of order $p$ where $p$ is the number of outputs.
Numerical examples demonstrate the usefulness of reported results where we consider unobservable and uncontrollable systems.			
\newpage
			
\end{document}